# Matching Researchers to Funding Calls: A Reproducible Institution-Level Framework

Wenceslao Arroyo-Machado[1,2*]; Laura Lázaro-Soraluce[2]; Clara Ortega-Sevilla[2]; Enrique de la Fuente-Gutiérrez[2]; & Daniel Torres-Salinas[2,3]

[1]INGENIO (CSIC-UPV), València, Spain
[2]EC3metrics Spin-Off, University of Granada, Granada, Spain
[3]Department of Information and Communication Sciences, University of Granada, Granada, Spain
*Corresponding author: wences@ingenio.upv.es

**Abstract**

Grant recommendation systems remain one of the least explored areas within academic recommender systems, and existing proposals are typically tied to specific funding agencies or disciplinary domains. This paper presents an institution-level reproducible framework for matching researchers to funding opportunities by combining bibliometric profiling with semantic matching. Rather than representing each researcher through a single aggregated profile, the framework constructs multiple publication sets defined by bibliometric criteria such as authorship position and time window, each independently compared against funding calls using word embeddings. Within-researcher normalisation and percentile-based ranking transform cosine similarity scores into actionable recommendations. A case study applied to 3,013 researchers from the University of Granada and 291 Horizon Europe topics verify it and shows that the four indicators capture complementary signals.

**Keywords**

Research funding; Grant recommendation; Bibliometric profiling; Semantic matching; SPECTER2; Horizon Europe; Scopus

**Contributorship statement**

WAM - Conceptualization; Investigation; Methodology; Project administration; Supervision; Visualization; Writing - Original Draft; Writing - Review & Editing.
LLS - Data Curation; Formal analysis; Investigation; Software; Writing - Original Draft.
COS - Data Curation; Formal analysis; Investigation; Software; Writing - Original Draft.
EdlFG - Funding acquisition; Project administration; Resources.
DTS - Validation.

# 1. Introduction

Science is not free. Research requires financial resources that go beyond institutional budgets, making external funding an essential component of academic activity (Thelwall et al., 2023). Despite the undeniable societal benefits of scientific investment, quantifying and communicating these returns in terms that the general public can readily understand remains a considerable challenge (Lane, 2009; Yin et al., 2022). Nevertheless, its direct impact and implications within the scientific realm is more tangible. Funding does not merely make scientific progress possible, but it also enhances research outputs, though its effect depends on the diversity of funding sources and rarely shows in the short term (Ding & Bu, 2025; Gök et al., 2016). Moreover, and beyond productivity, external funding also influences academic relationships, by raising international collaborations (Ubfal & Maffioli, 2011). However, despite such benefits, this dependence on external support is not without its complications. Securing and managing funding imposes significant administrative and bureaucratic demands on researchers. The rise in pressure to maintain financial continuity can gradually shift attention away from scientific inquiry itself and towards the perpetual pursuit of the next grant (Ioannidis, 2012).

Research funding comes from a multiplicity of actors operating at different levels and driven by their own rationales and interests (Buckley, 2022; Røttingen et al., 2013). Rather than a blank cheque, such funding comes with predefined thematic priorities that researchers must meet, or adapt to, in order to apply for successfully ('Don't Deprioritize Curiosity-Driven Research', 2026). This dependency on external criteria is not without consequences. At its most extreme, this can give rise to active conflicts of interest, as illustrated by the controversy surrounding the science of consciousness (IIT-Concerned et al., 2023). More pervasively, it contributes to a structural misalignment between investment priorities and actual needs (Kumar et al., 2024; Yegros-Yegros et al., 2020). Despite these risks, the funding landscape is broad and diverse enough that researchers will rarely struggle to find opportunities aligned with their interests. The challenge, however, has become navigating an increasingly competitive environment. In this sense, success rates for some European calls have fallen to single-digit figures (Naddaf, 2025), dropped from 26% to 19% for early-stage investigators at the NIH (Kaiser, 2026), and similarly declined in China following a surge in applications (Conroy, 2024). Keeping pace with this evolving landscape is a growing challenge, and success in research funding depends heavily on the active pursuit of opportunities (Bol et al., 2018).

From a scientometric perspective, the dynamics and implications of research funding constitute one of the field's recurring topics of inquiry, with acknowledgements in scholarly publications serving as the primary data source for its analysis (Álvarez-Bornstein & Montesi, 2021). Within this framework, several fronts of research have been developed. Efforts have been made to map funding thematically, either to characterise the portfolio of a specific agency (Belter, 2013) or to describe the global funding landscape within a particular field, such as oncology (Schmutz et al., 2019). Alongside this, frameworks have been proposed to account for the structural complexity of funding systems, recognising the multilevel and heterogeneous nature with which researchers combine funding sources in practice (Aagaard et al., 2021). Inequalities in funding access have also been documented, with biases linked to researchers' gender (Bol et al., 2022), institutional size and prestige (Murray et al., 2016), and research area (Tian et al., 2024). Equally, it has been examined the impact of funding on scientific output (Jacob & Lefgren, 2011), on academic careers (Bloch et al., 2014), and on the monitoring and evaluation mechanisms that funders themselves apply to the research they support (Abudu et al., 2022). However, most of these studies are descriptive and explanatory in nature, with little bearing on the application process itself.

One of the most direct and forward-looking applications has been the development of funding recommendation systems. However, grant recommender systems represent one of the least explored areas within academic recommender systems (Z. Zhang et al., 2023). Finding relevant funding in large databases is a difficult and time-consuming process for most researchers, and the commercial tools available offer very limited search capabilities (Zhu et al., 2023). Early systems addressed this problem through keywords and association rules (Kamada et al., 2015, 2016), whilst later proposals incorporated information retrieval techniques such as TF-IDF or BM25 (Acuna et al., 2022). More recent developments have incorporated language models such as BERT (Devlin et al., 2018), or specialised biomedical embeddings such as BioWordVec (Y. Zhang et al., 2019) and BioSentVec (Chen et al., 2019). However, despite this progress, most of these proposals share a structural limitation. They are *ad hoc* solutions, developed for specific domains, institutions, or funding sources, with limited capacity for generalisation and difficult to transfer to other contexts. Moreover, their bibliometric approach relies solely on metadata fields such as titles and abstracts to construct the semantic connection.

No grant recommendation system to date offers a scalable solution across different institutions that fully exploits bibliometric data. This paper addresses that gap by proposing a method for aligning research funding opportunities with researcher profiles through the combination of bibliometric and semantic techniques. This approach aims to support research institutions at two complementary levels. At the individual level, it alerts researchers to the funding opportunities most aligned with their profile, reducing the effort and uncertainty involved in funding searches. At the institutional level, it enables research offices to identify, for each call, the most suitable candidate within their community, providing a basis for the strategic allocation of research expertise across funding opportunities. The specific objectives are:
1. To design a methodological framework that integrates bibliometric indicators and semantic methods to measure the alignment between research funding calls and researcher profiles.
2. To validate the framework through a case study drawing on researcher profiles from the University of Granada (UGR) and *Horizon Europe* topics, using *Scopus* as the primary bibliographic data source.

The paper is structured as follows. Section 2 describes the proposed methodological framework, outlining its theoretical foundations and core components. Section 3 details the case study, including the methodology adopted, the data collected, and the indicators employed. Section 4 presents the results obtained from the case study application. Section 5 discusses the findings and examines the limitations of the framework. Finally, Section 6 draws the conclusions of this work.

## 2. Methodological framework

The methodological framework is presented below and summarised in **Figure 1**. The workflow is organised into three main blocks: (1) database generation, comprising bibliographic and authorship data alongside calls for proposals; (2) multidimensional researcher profiling, which prepares both researcher and call data for subsequent comparison and matching; and (3) scoring and ranking, which compares and identifies the most relevant funding opportunities for each researcher. It should be noted that, as a methodological framework, it is not designed to be bound to a specific dataset, context, or configuration, but to remain adaptable across multiple scenarios. Also, it is conceived to support research institutions in monitoring funding opportunities and aligning them with their researchers' profiles, with part of its potential residing in its ability to discriminate and identify the most suitable researchers within the institution. The case study, described subsequently, specifies a concrete setting and details the technical implementation of the framework in practice.

**Figure 1**. Methodological framework for matching researchers to funding opportunities.

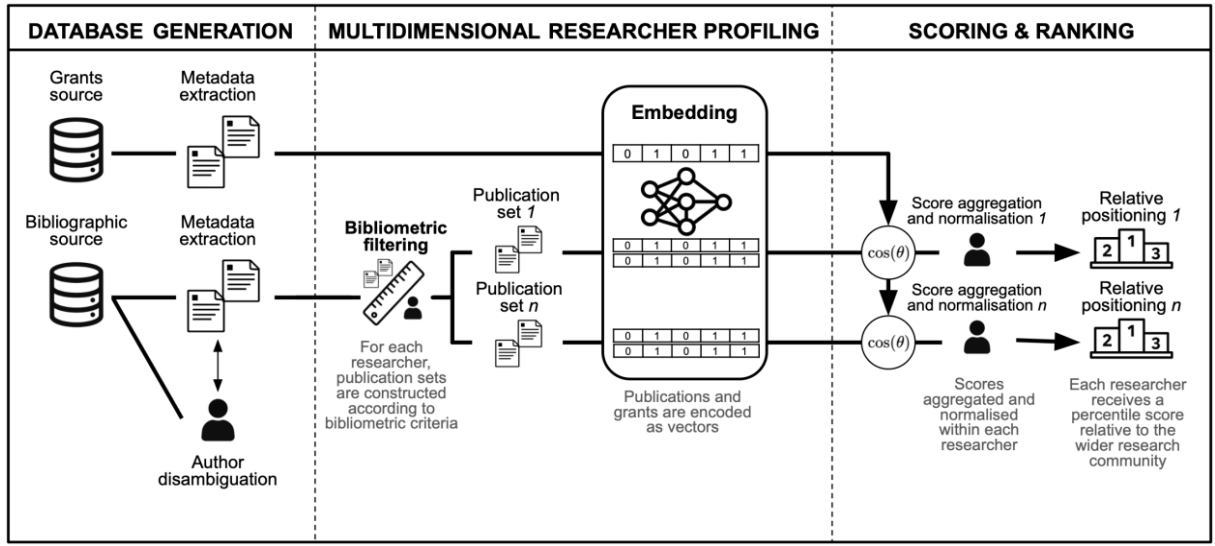

## 2.1. Database generation

### 2.1.1. Data extraction

The first step is the construction of the database, which draws on at least two independent sources. On one hand, a bibliographic source providing publication records and authorship data, together comprising sufficient metadata to support basic bibliometric operations. On the other, a funding source providing basic information on the calls for proposals.

The bibliographic source may consist of a single database or a combination of several, to obtain broader coverage and a more comprehensive understanding of the institution's research output. This may be particularly relevant for institutions with a more multidisciplinary profile or whose publications are not exclusively oriented towards international journals, for which databases such as *Web of Science* or *Scopus* may offer biased coverage (Asubiaro et al., 2024; Gusenbauer, 2024). Where available, a Current Research Information System (CRIS) may itself serve as the primary source, as it is specifically designed to capture the full scope of an institution's research activity and typically holds normalised researcher records alongside production tracked across multiple sources. In any case, the selected source or combination of sources must meet a set of minimum requirements:

1. It must provide coverage that is sufficiently representative of the institution's output over a minimum period to constitute a reliable baseline.
2. Since it will underpin the semantic matching process, it is necessary to ensure that as many publications as possible include, in addition to the title, an abstract; a publication lacking this will not be excluded but will be processed with less information. This is a pertinent consideration when selecting a source, as some publishers have ceased to make abstracts openly available, thereby affecting open sources such as *Crossref* and *OpenAlex* ('Publishers Close Access to Scholarly Content Such as Abstracts Due to AI Incentives', 2025). The presence of keywords or topics (e.g., *OpenAlex* topic classification) enriches the representation when available, but their absence does not compromise the process in the same way.
3. The source must include some mechanism for author identification, whether a persistent identifier (e.g., *ORCID*, *Web of Science ResearcherID*, *Scopus Author ID*…), email

   address, or a normalised name, to facilitate the unambiguous aggregation of each researcher's output and to verify that the works assigned to a given individual genuinely belong to them.
4. The source must provide the metadata fields required by the bibliometric criteria to be applied. For instance, if the framework is configured to filter by publication year or authorship position, the source must include reliable information for those fields.

The funding source, for its part, provides the counterpart to the bibliographic data, with less demanding requirements. Just as publications require a title, abstract and keywords for embedding vectorisation, analogous information is needed here. Apart from the title, calls for proposals do not typically include a summary that meets the formal characteristics of a scientific abstract. However, they must offer details on the thematic scope, lines of work or expected outcomes that serve this purpose and allow the focus of the call to be properly understood. Keywords may likewise appear in the form of a controlled thesaurus or a thematic classification (e.g., MeSH, EuroSciVoc, UNESCO codes). The aim is to construct, from the available information, a document structurally analogous to a scientific publication, as summarised in **Table 1**. Ultimately, both publications and calls must be representable through the same set of textual fields, even if their nature and form differ, sharing ultimately a common academic language.

**Table 1**. Field equivalence between publications and calls for proposals.

| Field | Publications | Calls | Description |
|---|---|---|---|
| *Title** | Title | Title | Concise identifier of the document's subject matter |
| *Abstract* | Abstract | Description<br>Destination<br>Expected outcome<br>Scope | Core textual field for semantic vectorisation |
| *Keywords* | Keywords<br>Topics | Keywords<br>Category<br>Classification<br>Thesaurus | Controlled or free-text terms that reinforce the thematic signal |

*Mandatory in both cases as the minimum unit of semantic content

### 2.1.2. Researcher data normalization

Of the two sources described above, the bibliographic one(s) requires a processing step before any further analysis can be carried out. A single researcher may appear under multiple profiles in a bibliographic database, and correctly merging and assigning their publications is the foundation for the bibliometric layer of the framework, from which their multidimensional profile will subsequently be constructed. This process is facilitated when the source provides information such as *ORCID*, email addresses, or institutional affiliations. Regardless of the database used, this step is always necessary, although the effort involved will vary depending on the quality and normalisation of the source data (Chinchilla-Rodríguez et al., 2024). However, as noted above, where a CRIS serves as the primary source, this step is considerably simplified, as researcher profiles are typically already resolved.

## 2.2. Multidimensional researcher profiling

*2.2.1. Bibliometric profiling*

Prior approaches to researcher-grant matching typically represent researchers through aggregated profiles or researcher-defined keyword queries (Z. Zhang et al., 2024; Zhu et al., 2023), or cluster publications into thematic groups before comparison (Zhu et al., 2023). The present framework departs from this logic. Rather than treating the researcher as a single entity, it constructs multiple publication sets per researcher, each defined by a specific bibliometric criterion that isolates a distinct signal about their role, interests or thematic orientation. Each set is compared independently against the call, producing not a single score but a set of indicators that together profile the researcher's alignment, closer to a dashboard than to a summary measure.

The criteria used to define these sets can vary depending on analytical goals and available data. Authorship position is one option, as first, last or corresponding authorship conventionally signals a more central intellectual contribution or leadership role (Chinchilla-Rodríguez et al., 2019; Rennie, 2000). Others might include publications above a given impact threshold, works involving international collaboration, or output within a specific subfield. Each criterion isolates a subset of the researcher's output that constitutes interpretable evidence about their expertise and interests. To ensure scores rest on a sufficient empirical base, minimum publication thresholds are applied. As a result, researchers who do not meet the threshold for a given set are excluded from that ranking but may still appear in others.

The time frame operates as a cross-cutting dimension that can be applied in combination with any of the criteria above. Rather than defining a different type of publication, it modulates the time window over which each set is constructed, so that the same criterion can be applied over a medium-term horizon or over a shorter window, which is more sensitive to recent shifts in focus or the emergence of new research lines. Combining both horizons with each criterion allows the framework to distinguish between what a researcher has worked on and what they are working on now, a distinction that is particularly relevant when the goal is to identify candidates with both depth of expertise and current engagement with a given topic.

Furthermore, defining publication sets prior to vectorisation also offers a computational advantage, as only the publications belonging to a given set need to be encoded. Publications that do not meet any criterion are not vectorised at all, reducing the overall computational cost.

*2.2.2. Semantic matching*

Early grant recommendation systems relied exclusively on lexical representations, such as keyword matching and association rules (Kamada et al., 2015, 2016), which assume shared vocabulary between researchers and calls. As computational capacity and machine learning techniques advanced, more sophisticated approaches emerged, with word embeddings now outperforming these lexical methods in grant recommendation tasks (Z. Zhang et al., 2024; Zhu et al., 2023). Each publication within a publication set, alongside the call, is transformed into such a vector using the textual fields described in Section 2.1.1. (title, abstract and keywords where available, and their equivalent for funding calls). This vectorisation is performed at the individual publication level, rather than at the level of an aggregated researcher profile, which is precisely the operation for which these specific pre-trained models are designed and at which they perform most reliably.

Although general-purpose language models can be applied, the vocabulary, syntax and semantic conventions of academic publications differ substantially from general text, and models trained on broad corpora tend to underperform when applied to domain-specific matching tasks (Fahrudin et al., 2025; Wolff et al., 2024). Specialised models encode the way in which scientific concepts relate to one another, producing more coherent and discriminating similarity scores when comparing publications to funding calls. This advantage is especially relevant for multidisciplinary institutions, where the thematic breadth of the research portfolio requires a model capable of representing diverse scientific domains with comparable fidelity. Beyond model selection, further refinement is possible through preprocessing and representation strategies, such as text normalization and field-level weighting, which can improve the discriminatory power of the resulting vectors.

There are several embedding models for scientific text representation. **Table 2** summarises the most widely discussed options, all trained on academic corpora and therefore suitable candidates for this type of application. Empirical comparisons across classification and similarity tasks consistently show that models pre-trained on scientific literature outperform general-purpose alternatives, with *SPECTER2* and *SciBERT* emerging as the most robust options across disciplines (Fahrudin et al., 2025; Wolff et al., 2024). Domain-specific models such as *BioSentVec* or *BioWordVec* may offer additional precision in institutions with a predominantly biomedical focus (Z. Zhang et al., 2024; Zhu et al., 2023), though their advantage tends to diminish outside that domain. The choice of model should therefore be guided by the disciplinary profile of the institution and the coverage of the available training corpus relative to its research output.

**Table 2**. Main pre-trained word embedding models for academic text.

| Model | Training corpus | Base model | Model type | Research areas |
|---|---|---|---|---|
| *SPECTER2* (Singh et al., 2022) | ~6M citation triplets | SciBERT | Transformer | Multidisciplinary |
| *SciNCL* (Ostendorff et al., 2022) | S2ORC | BERT | Transformer | Multidisciplinary |
| *SciBERT* (Beltagy et al., 2019) | Semantic Scholar | BERT | Transformer | Biomedical and computer science |
| *BioSentVec* (Chen et al., 2019) | PubMed + MIMIC-III clinical notes | — | sent2vec | Biomedical science |
| *BioWordVec* (Y. Zhang et al., 2019) | PubMed + MeSH | — | fastText | Biomedical science |

## 2.3. Scoring and ranking

With publication and call vectors in place, the next step is to quantify their semantic proximity. This is done using cosine similarity, which measures the angle between two vectors in the embedding space and is independent of document length:

$$\text{sim}(p, c) = \frac{\vec{p} \cdot \vec{c}}{|\vec{p}| \cdot |\vec{c}|}$$

where $\vec{p}$ and $\vec{c}$ represent the embedding vectors of a publication and a call respectively, and the resulting score ranges from 0 to 1, with higher values indicating greater semantic proximity.

Once cosine similarity scores have been computed for each publication-call pair within a given publication set, these scores must be aggregated into a single value per researcher and set. The default aggregation is the arithmetic mean of all similarity scores in the set, which provides a balanced estimate of the overall thematic alignment between a researcher's output and the call. However, when a researcher has a large and thematically diverse number of publication records, the mean can be pulled downwards by works that are only loosely related to the call. In such cases, the aggregation is performed over the top 33% of publications by similarity score, capturing the most aligned third of the researcher's output and providing a more discriminating estimate of their peak thematic relevance.

After that, and prior to ranking, each researcher's aggregated similarity scores are standardised using a z-score transformation applied across all their publication-call pairs within a given set. This normalisation step rescales individual scores relative to each researcher's own distribution, rather than across the full population. As a result, researchers whose publication sets span a broad range of themes receive lower effective scores on any single call, whilst those with a more focused profile are rewarded for their thematic concentration. The z-score is computed as:

$$z_{s,r} = \frac{a_{s,r} - \mu_r}{\sigma_r}$$

where $a_{s,r}$ is the aggregated similarity score of researcher $r$ for publication set $s$, and $\mu_r$ and $\sigma_r$ are the mean and standard deviation of all aggregated scores for that researcher across all publication sets.

The aggregated and normalised score is a continuous value that, taken in isolation, lacks direct interpretability. A score for a given researcher on a given call carries little meaning without knowing whether it is high or low relative to the rest of the community. The framework therefore positions each researcher's score within the distribution of scores obtained by all researchers in the institution for that same call and publication set. This relative positioning is expressed as a percentile rank, so that researchers in the upper percentiles of the distribution are identified as the most promising candidates for a given call. This step transforms an abstract score into an actionable recommendation, grounded in the comparative performance of the full research community rather than in any absolute threshold. Beyond individual recommendations, ranking researchers relative to their own community transforms the framework into an instrument of institutional strategy, as for each call, it identifies the best available match within the institution, naturally distributing visibility across different research profiles.

## 3. Case study
### 3.1. Data and sources
The framework was implemented and evaluated using the University of Granada (UGR) as a case study, one of the largest public research universities in Spain, with over 5,487 researchers affiliated across 123 departments at the time of analysis in January 2026[1]. The primary bibliographic source was *Scopus*, supplemented by *OpenAlex* for thematic enrichment. In this case study, the availability of the UGR's Current Research Information System (CRIS)[2] provided a curated master list containing each researcher's verified *Scopus* identifier, *ORCID* iD and institutional email address. Researchers present in Scopus but absent from the CRIS were excluded from the study population. This made the CRIS the reference for population definition and enabled the detection and merging of fragmented Scopus author profiles.

Bibliographic data were retrieved from *Scopus* through a two-stage process. In the first stage, all UGR-affiliated records (limited to article, review, letter, book and book chapter) from the last five years (2021-2025) were downloaded via both a manual export and the *Scopus* API. The manual export was necessary to capture the correspondence address field with the email, which the API does not provide due to privacy reasons. This allowed the framework to pair corresponding author email addresses with their *Scopus* Author IDs, with an additional verification step to ensure the email was assigned by Scopus to the correct author. The API download served to retrieve full publication metadata and author-level information. Author records were then unified with the CRIS data through a cascading matching procedure applied sequentially by *Scopus* Author ID, *ORCID*, email address and, as a last resort, normalised name string. Where two records shared the same verified name and at least one email address, they were merged into a single profile, consolidating all associated information.

In the second stage, once the author database had been consolidated, the *Scopus* API was queried again using each researcher's verified *Scopus* Author ID to retrieve all their publications from the study period, regardless of institutional affiliation. This step ensured that output produced prior to or outside the UGR was also captured. The resulting bibliographic database comprised 30,077 publications and 3,759 *Scopus* author profiles, corresponding to 3,013 researchers registered in the CRIS with at least three publications. This minimum defines the potential population for analysis, regardless of any additional filters applied subsequently, and represents 55% of the total institutional population. Thematic enrichment was carried out via the *OpenAlex* API, from which the primary topic was retrieved for each publication by DOI.

The call dataset consisted of 291 *Horizon Europe* topics (hereafter referred to as calls) scraped from the *European Commission's Funding and Tenders Portal*[3], covering all six clusters open as of September 2025. For each call, the title and topic description were extracted and combined to construct a document equivalent in structure to a scientific abstract

### 3.2. Data processing and methods
Researcher-call alignment was assessed through four indicators, constructed by crossing two temporal windows with two publication set criteria and summarised in **Table 3**. The temporal dimension distinguishes between a five-year horizon, which captures consolidated thematic expertise, and a two-year window, which is more sensitive to recent shifts in research focus. The rationale is that a researcher may have broadened or redirected their agenda over time, and recent output is not always representative of their full track record, nor vice versa. The publication set dimension distinguishes between a researcher's full output and a subset restricted to first, last or corresponding author positions, which conventionally signals a more central intellectual contribution, separating works the researcher has actively led from those where their role was more peripheral. Together, the four indicators offer complementary rather than redundant signals, allowing the framework to distinguish between depth of expertise and current engagement.

---

[1] https://produccioncientifica.ugr.es/grupos
[2] https://produccioncientifica.ugr.es/
[3] https://ec.europa.eu/info/funding-tenders/opportunities/portal/

**Table 3**. Multidimensional indicators used to assess researcher-call alignment.

| Indicator | Description | Publication set | Time window | Threshold |
|---|---|---|---|---|
| *Research background* | Measures the researcher's consolidated thematic alignment with the call | All publications | 5 years | >=5 publications |
| *Current focus* | Captures the researcher's recent thematic orientation | All publications | 2 years | >=3 publications |
| *Research leadership* | Reflects thematic alignment based solely on publications where the researcher led the research | First, last or corresponding author only | 5 years | >=4 publications |
| *Current leadership* | Identifies researchers whose most recent output, restricted to publications where they led the research, aligns with the call | First, last or corresponding author only | 2 years | >=2 publications |

Minimum publication thresholds were defined heuristically through iterative testing. Initial values were evaluated by inspecting borderline cases at both extremes, that is, thresholds set too high excluded researchers with relevant profiles but a limited number of publications, while thresholds set too low admitted profiles with insufficient empirical basis or tangential relationship. Manual inspection of these edge cases was instrumental in calibrating the final values. For the *Research background* and *Current focus* sets, the minimum required output is five and three publications respectively. For the *Research leadership* and *Current Leadership* sets, the thresholds are four and two publications. Researchers who do not meet the threshold for a given set are excluded from the corresponding ranking but may still appear in others. It should be noted that these thresholds do not constitute a universal recommendation. They represent a pragmatic calibration decision for the specific case study presented here, and their appropriateness will vary with the size and characteristics of the research community under analysis. Conceptually, a threshold of this kind could even be operationalised as a soft boundary, applying a penalty or weight reduction to researchers whose publication count falls just below the cut-off, rather than enforcing a strict binary exclusion.

All publications and calls were vectorised using *SPECTER2*, selected on the basis of its consistent performance across disciplines in prior benchmarking studies (Wolff et al., 2024). It is important to note that cosine similarity scores produced by *SPECTER2* tend to cluster at high values[4], reducing the ability to discriminate between more and less relevant matches (Dhakal et al., 2025). To address this, a common-component removal procedure was applied. Orthogonal projection, a common debiasing technique in machine learning and NLP, was applied to remove the stylistic bias introduced by the academic writing conventions shared across all texts. An empty document following the input structure was encoded by *SPECTER2*, producing a vector that captures this generic style rather than any specific content. Each document vector was then projected onto the orthogonal complement of this baseline, so that the resulting representation reflects thematic content alone.

---

[4] https://github.com/allenai/SPECTER2/issues/14

Then, cosine similarity was computed between each publication vector and each call vector. The resulting per-publication scores were then aggregated into a single researcher-level score for each publication set. The aggregation rule was determined heuristically through iterative testing, inspecting cases where extreme values introduced noise or where overly restrictive thresholds excluded relevant profiles. The final rule applies the arithmetic mean when the publication set contains few works, and switches to the mean over the top 33% of publications by similarity score when the set is sufficiently large. Specifically, if the top 33% of a researcher's publications in a given set amounts to two papers or fewer (less than 7 publication records), the arithmetic mean over the full set is used instead; otherwise, the top 33% mean is applied. This prevents individual highly aligned but isolated works from disproportionately inflating the score of researchers whose overall output is not genuinely close to the call.

Prior to aggregation, each researcher's similarity scores were normalised within their own distribution, rescaling individual scores relative to each researcher's profile rather than across the full population. The aggregated and normalised scores were then used to compute percentile ranks across the full institutional population for each call and publication set. For the purposes of this analysis, researchers at or above the 95th percentile were treated as strong candidates, as this threshold identifies the top 5% of the community by alignment with a given call.

### 3.3. Data and code availability
All scripts were developed in *R* and *Python*. *R* was used for database construction, including *Scopus* data retrieval, bibliometric processing and indicator construction. *Python* handled the embedding and similarity pipeline, covering vectorisation and cosine similarity computation, with the sentence-transformers library providing the *SPECTER2* implementation. The embedding model was executed on *Google Colaboratory*. All scripts are fully replicable and openly available at GitHub (https://github.com/claraaortega/Paper-Call-Project).

## 4. Results
The results are presented in two complementary analyses. The first examines how researcher assignments are distributed across calls and indicators, assessing whether recommendations spread broadly or concentrate on a few individuals, and whether the four indicators provide overlapping or distinct outputs. The second explores a specific case in depth to provide a qualitative validation of the recommendations produced.

### 4.1. Assignment distribution and indicator complementarity
Of the UGR researchers identified in *Scopus*, 2,662 (88%) appear in at least one of the 291 calls with an assignment under at least one indicator (**Table 4**). Coverage is broad, yet it varies considerably across indicators. *Research background* produces the highest number of assignments (2,540 researchers), whilst *Current leadership*, which applies the most selective criteria, covers 1,432. Each indicator also captures researchers that the others miss. *Research background* returns 297 unique researchers not identified by any other indicator, *Research leadership* adds 55, and *Current focus* contributes 24. *Current leadership* adds no exclusive assignments, meaning that every researcher it identifies is also captured by at least one other indicator. Across all indicators combined, researchers receive an average of 20.1 unique call recommendations. Around half (1,281) receive recommendations under all four indicators, whilst 376 (14%) appear under only one.

**Table 4**. Assignment summary by indicator

|  | Researchers | Unique researchers | Average call per researcher | Average researcher per call |
|---|---|---|---|---|
| *Research background* | 2,540 | 297 | 14.6 | 128.0 |
| *Current focus* | 2,004 | 24 | 14.6 | 101.2 |
| *Research leadership* | 1,883 | 55 | 14.6 | 95.0 |
| *Current leadership* | 1,432 | 0 | 14.7 | 72.4 |
| **Total** | 2,662 | --- | 20.1 | 184.3 |

**Figure 2** shows the distribution of recommended calls per researcher for each indicator. All four distributions are right-skewed and strikingly similar in shape, with a median of 13 calls per researcher and comparable interquartile ranges. This similarity, however, does not imply redundancy. It is a direct consequence of the shared design logic. Each indicator applies the same 95th percentile threshold independently for every call, which mechanically generates a comparable number of assignments per call regardless of the indicator used. The average number of recommended calls per researcher is therefore approximately 14 for each indicator individually, rising to 20.1 when unique calls across all indicators are combined. In other words, the distributions look alike because the selection mechanism is structurally identical, not because the indicators recommend the same people.

**Figure 2**. Distribution of recommended calls per researcher for each indicator. Left panels show histograms; the right panel displays the corresponding box plots.

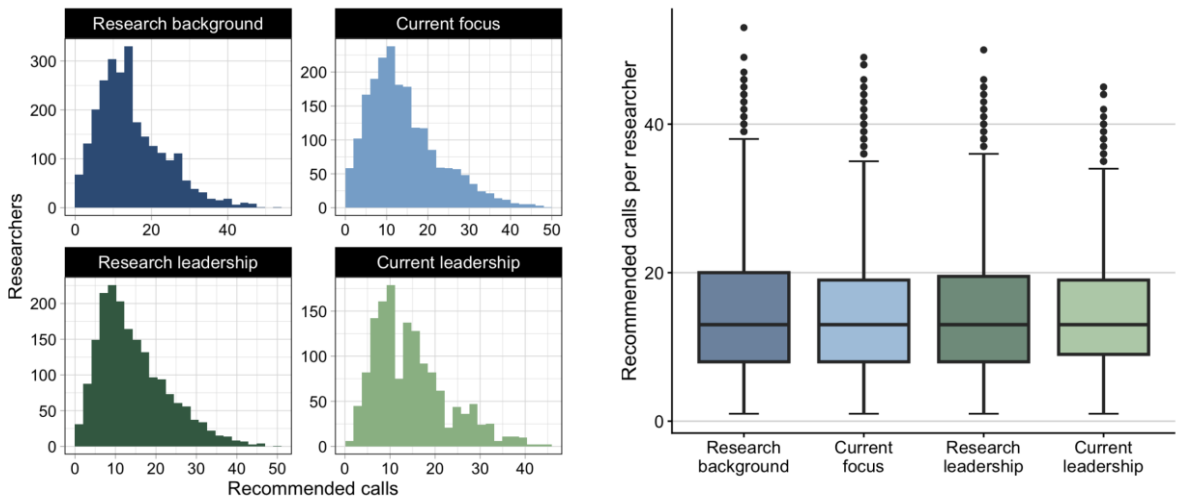

To assess whether the indicators identify the same researchers and rank them similarly, assignment overlap and percentile rank correlation were examined jointly (**Table 5**). Overlap values range from 37% to 75%, and Spearman rank correlations between 0.55 and 0.71, confirming that no two indicators are interchangeable. The clearest distinction runs between the general indicators (*Research background* and *Current focus*) and the leadership-based ones (*Research leadership* and *Current leadership*). Within each group, assignments and rankings are more consistent, but, across groups, both overlap and correlation drop noticeably, with cross-pair values reaching as low as 37% and ρ = 0.55. The asymmetry in overlap values is also informative. Leadership indicators tends to show higher overlap into the general ones (65-75%), but this reflects a size effect rather than redundancy. Because leadership indicators produce fewer assignments, a larger share of their pairs naturally satisfies the less restrictive thresholds of the general indicators. The reverse direction confirms this interpretation. Only 37-56% of general-indicator pairs also appear in the leadership indicators, indicating that the broader indicators capture a substantial set of researchers whom the leadership criteria exclude.

**Table 5**. Pairwise overlap between indicators, expressed as the percentage of call-researcher pairs in the row indicator also present in the column indicator, and the Spearman correlation between the percentile ranks assigned by each indicator pair.

| Call–researcher pairs and percentile correlation in row indicator (%) also present in column indicator | *Research background* | *Current focus* | *Research leadership* | *Current leadership* |
|---|---|---|---|---|
| ***Research background*** |  | 58% ρ=0.69 | 56% ρ=0.71 | 37% ρ=0.56 |
| ***Current focus*** | 74% ρ=0.69 |  | 51% ρ=0.55 | 53% ρ=0.67 |
| ***Research leadership*** | 75% ρ=0.71 | 54% ρ=0.55 |  | 50% ρ=0.64 |
| ***Current leadership*** | 65% ρ=0.56 | 74% ρ=0.67 | 66% ρ=0.64 |  |

**4.2. Case study**

To assess whether the framework produces meaningful recommendations in practice, **Table 6** presents the two highest-ranked calls for a researcher affiliated with the Department of Criminal Law at UGR, whose work centres on criminal law and criminology, with particular emphasis on cybercrime and the protection of fundamental rights. In recent years, this researcher's output has increasingly engaged with artificial intelligence and big data, mirroring a broader trend across the social sciences and humanities. Across all four indicators, the framework recommends a total of 38 unique calls for this researcher. **Table 6** focuses on the two highest-ranked calls per indicator to illustrate how the prioritisation shifts depending on the dimension considered. Seven of the eight entries belong to *Horizon Europe* Cluster 3 (civil security for society) and address border security, customs and supply chain integrity, organised crime prevention and operational cybersecurity. The exception is a Cluster 2 call (culture, creativity and inclusive society) on advisory support to counter disinformation and foreign information manipulation and interference (FIMI), which explicitly targets the application of AI and big data to detect and debunk disinformation. All eight entries are thematically coherent with the researcher's documented expertise, and none can be dismissed as a false positive or an artefact of the matching process.

Most of these calls appear across multiple indicators, yet the ranking they receive differs substantially. The call on customs and supply chain security (HORIZON-CL3-2025-01-BM-03), for instance, features in three of the four top-two lists but occupies a different position in each, reflecting how each indicator weighs the researcher's profile differently. Under *Current leadership*, it is ranked first with a percentile of 100, meaning that the researcher's recent leading-author output is the most closely aligned in the entire institutional population. More revealing is the behaviour of *Current focus*, which is the only indicator to place the FIMI call among the top two, ranking it in first position (2nd rank, 99.95th percentile). This result directly captures the researcher's recent pivot towards AI-related topics. Because *Current focus* operates on a two-year window, it is sensitive to thematic shifts that the five-year indicators inevitably dilute. Had the framework relied solely on *Research background*, this highly relevant call would not have appeared among the top recommendations. Conversely, *Research leadership* uniquely surfaces the call on organised crime intelligence (HORIZON-CL3-2025-01-FCT-03) in its top two, reflecting the researcher's longer-standing expertise as a leading author in that area. The case confirms that the four indicators, rather than merely expanding coverage, reshape the priority order in ways that carry practical significance for the researcher.

**Table 6.** Top-two recommended calls per indicator for the selected researcher.

| Indicator | Call | Rank | Percentile |
|---|---|---|---|
| *Research background* | HORIZON-CL3-2025-01-BM-02<br>Open topic on secured and facilitated crossing of external borders | 2nd | 99.96 |
| | HORIZON-CL3-2025-01-BM-03<br>Open topic on better customs and supply chain security | 2nd | 99.96 |
| *Current focus* | HORIZON-CL2-2025-01-DEMOCRACY-01<br>Advisory support and network to counter disinformation and foreign information manipulation and interference (FIMI) | 2nd | 99.95 |
| | HORIZON-CL3-2025-01-BM-03<br>Open topic on better customs and supply chain security. | 5th | 99.80 |
| *Research leadership* | HORIZON-CL3-2025-01-FCT-03<br>Open topic on improved intelligence picture and enhanced prevention, detection and deterrence of various forms of organised crime | 2nd | 99.95 |
| | HORIZON-CL3-2025-02-CS-ECCC-02<br>New advanced tools and processes for Operational Cybersecurity | 2nd | 99.95 |
| *Current leadership* | HORIZON-CL3-2025-01-BM-03<br>Open topic on better customs and supply chain security | 1st | 100 |
| | HORIZON-CL3-2025-02-CS-ECCC-02<br>New advanced tools and processes for Operational Cybersecurity | 1st | 100 |

## 5. Discussion

This work presents the first grant recommendation framework that combines bibliometric profiling with semantic matching in a reproducible and scalable architecture. Previous systems have been designed for specific funding agencies and disciplinary domains. Zhu et al. (2023) built a BERT-based classifier trained on *NIH* publication-grant pairs, whilst Z. Zhang et al. (2024) employed biomedical word embeddings over the same ecosystem. Both produce effective recommendations within their scope, but their architectures are tightly coupled to *PubMed* and *NIH* data, requiring retraining to operate elsewhere. The present framework separates the methodological logic from its data instantiation, so that any combination of bibliographic source, embedding model and funding programme can be plugged in without reengineering the pipeline. The case study, applied to the University of Granada and 291

*Horizon Europe* topics, shows that recommendations reach a broad share of the research community rather than concentrating on a few dominant profiles.

A second contribution is the multidimensional profiling strategy. Prior approaches represent researchers either through a single aggregated profile (Z. Zhang et al., 2024) or through unsupervised thematic clusters (Zhu et al., 2023), yielding one recommendation list per researcher. The present framework constructs multiple publication sets per researcher, each defined by a distinct bibliometric criterion, and compares each independently against the call. The resulting indicators distinguish consolidated expertise from recent activity, and leading contributions from peripheral ones. This granularity is particularly relevant for research offices, which need to assess not only thematic proximity but also the nature of a candidate's engagement with a topic.

The within-researcher normalisation deserves specific attention. By standardising similarity scores relative to each researcher's own distribution, the system penalises thematically dispersed profiles and rewards concentrated ones. An important consequence is that scores are recalculated with every new call, making the framework inherently dynamic. Rankings shift as both the researcher's output and the available calls evolve, so no score accumulates over time, that is, each ranking reflects alignment with a specific opportunity at a specific moment. The same individual may therefore rank very differently depending on the moment, but this variability captures the reality that relevance is always relative to the funding opportunity at hand.

The framework also opens practical avenues beyond individual recommendation. By cross-referencing indicators, it becomes possible to detect complementarities within the institution, pairing a senior researcher with strong background alignment with an early-career colleague who scores well on current leadership, or identifying profiles from different departments that together cover the scope of a multidisciplinary call. Although this capability is not formalised in the present implementation, the underlying data structure supports it. Similarly, the fixed thresholds used in the case study could be replaced by adjustable controls in an institutional deployment, allowing research officers to modify parameters and recalculate rankings interactively.

## 5.1. Limitations

The principal methodological limitation is the absence of a ground truth dataset. Prior work has exploited retrospective *NIH* data linking funded outputs to specific calls at the researcher level (Z. Zhang et al., 2024; Zhu et al., 2023), sometimes complemented by expert ratings (Z. Zhang et al., 2024). *Horizon Europe* publishes funded project data through *CORDIS*, but only at the institutional level, without identifying which researcher led a given proposal. An alternative strategy based on publication acknowledgements is equally problematic, since the relationship between a researcher's bibliometric record and the calls they actually pursued is mediated by consortium composition, strategic decisions and institutional politics, none of which bibliometric data capture. Moreover, the multi-indicator design complicates quantitative benchmarking, as a researcher may rank differently depending on the time window or publication set considered. This is a deliberate choice rather than a shortcoming. Each indicator carries a distinct interpretive logic, and the traceability of individual publications behind each score supports informed human judgement rather than replacing it.

The reliance on *Scopus* in the case study limits the representativeness of the humanities and social sciences, whose output is often published through more local and specific channels. The framework itself is not bound to *Scopus*, and an institution seeking broader coverage could substitute or complement it with other sources as described in Section 2.1.1.

Some calls are inherently interdisciplinary, drawing on vocabularies that rarely co-occur in a single publication record. No individual researcher is likely to achieve high alignment with such calls unless their profile already bridges the relevant domains. The framework mitigates this through its multi-indicator structure, which can reveal partial alignments that a single score would obscure, but it does not fully resolve the challenge.

Finally, the percentile-based ranking introduces a structural tension. The system always identifies a top 5% for every call, even when no researcher has a meaningful thematic connection, potentially overstating the quality of the best available match. Conversely, for calls with many strongly aligned researchers, relevant candidates may fall just below the threshold. Users should therefore interpret percentile positions as relative standings within the institutional community rather than as absolute measures of suitability.

## 6. Final remarks

This paper has presented a reproducible framework for matching researchers to funding calls through the combination of bibliometric profiling with semantic matching. Unlike previous systems, which are bound to specific funding agencies and domains, the proposed approach separates the methodological logic from its data instantiation, allowing any bibliographic source, embedding model or funding programme to be integrated without re-engineering the pipeline. The case study confirms the practical viability of the approach, with the four indicators producing complementary rather than redundant recommendations, and qualitative inspection validating the coherence of individual profiles. The framework is conceived as an institutional tool that positions each researcher within their own community, supporting research offices in the strategic allocation of expertise to funding opportunities. Its reliance on openly available models and standard bibliometric data keeps computational costs low and makes the approach scalable to institutions of varying size and disciplinary breadth. Future work should focus on constructing ground-truth benchmarks for formal evaluation, extending the system to detect complementarities between researchers for multidisciplinary calls, and integrating it into an interactive platform for operational deployment.

## References

Aagaard, K., Mongeon, P., Ramos-Vielba, I., & Thomas, D. A. (2021). Getting to the bottom of research funding: Acknowledging the complexity of funding dynamics. *PLOS ONE*, *16*(5), e0251488. https://doi.org/10.1371/journal.pone.0251488

Abudu, R., Oliver, K., & Boaz, A. (2022). What funders are doing to assess the impact of their investments in health and biomedical research. *Health Research Policy and Systems*, *20*(1), 88. https://doi.org/10.1186/s12961-022-00888-1

Acuna, D. E., Nagre, K., & Matnani, P. (2022). *EILEEN: A recommendation system for scientific publications and grants* (arXiv:2110.09663). arXiv. https://doi.org/10.48550/arXiv.2110.09663

Álvarez-Bornstein, B., & Montesi, M. (2021). Funding acknowledgements in scientific publications: A literature review. *Research Evaluation*, *29*(4), 469–488. https://doi.org/10.1093/reseval/rvaa038

Asubiaro, T., Onaolapo, S., & Mills, D. (2024). Regional disparities in Web of Science and Scopus journal coverage. *Scientometrics*, *129*(3), 1469–1491. https://doi.org/10.1007/s11192-024-04948-x

Beltagy, I., Lo, K., & Cohan, A. (2019). SciBERT: A Pretrained Language Model for Scientific Text. *Proceedings of the 2019 Conference on Empirical Methods in Natural Language Processing and the 9th International Joint Conference on Natural Language Processing (EMNLP-IJCNLP)*, 3613–3618. https://doi.org/10.18653/v1/D19-1371

Belter, C. W. (2013). A bibliometric analysis of NOAA's Office of Ocean Exploration and Research. *Scientometrics*, *95*(2), 629–644. https://doi.org/10.1007/s11192-012-0836-0

Bloch, C., Sørensen, M. P., Graversen, E. K., Schneider, J. W., Schmidt, E. K., Aagaard, K., & Mejlgaard, N. (2014). Developing a methodology to assess the impact of research grant funding: A mixed methods approach. *Evaluation and Program Planning*, *43*, 105–117. https://doi.org/10.1016/j.evalprogplan.2013.12.005

Bol, T., De Vaan, M., & Van De Rijt, A. (2018). The Matthew effect in science funding. *Proceedings of the National Academy of Sciences*, *115*(19), 4887–4890. https://doi.org/10.1073/pnas.1719557115

Bol, T., De Vaan, M., & Van De Rijt, A. (2022). Gender-equal funding rates conceal unequal evaluations. *Research Policy*, *51*(1), 104399. https://doi.org/10.1016/j.respol.2021.104399

Buckley, R. C. (2022). Stakeholder controls and conflicts in research funding and publication. *PLOS ONE*, *17*(3), e0264865. https://doi.org/10.1371/journal.pone.0264865

Chen, Q., Peng, Y., & Lu, Z. (2019). BioSentVec: Creating sentence embeddings for biomedical texts. *2019 IEEE International Conference on Healthcare Informatics (ICHI)*, 1–5. https://doi.org/10.1109/ICHI.2019.8904728

Chinchilla-Rodríguez, Z., Costas, R., Robinson-García, N., & Larivière, V. (2024). Examining the quality of the corresponding authorship field in Web of Science and Scopus. *Quantitative Science Studies*, *5*(1), 76–97. https://doi.org/10.1162/qss_a_00288

Chinchilla-Rodríguez, Z., Sugimoto, C. R., & Larivière, V. (2019). Follow the leader: On the relationship between leadership and scholarly impact in international collaborations. *PLOS ONE*, *14*(6), e0218309. https://doi.org/10.1371/journal.pone.0218309

Conroy, G. (2024, November 14). Chinese scientists say funding shake-up has made it harder to win grants. *Nature*, *635*(8038), 270–270. https://doi.org/10.1038/d41586-024-03207-6

Devlin, J., Chang, M.-W., Lee, K., & Toutanova, K. (2018). *BERT: Pre-training of Deep Bidirectional Transformers for Language Understanding* (Version 2). arXiv. https://doi.org/10.48550/ARXIV.1810.04805

Dhakal, A., Paudel, K., & Sigdel, S. (2025). *An Artificial Intelligence Driven Semantic Similarity-Based Pipeline for Rapid Literature* (arXiv:2509.15292). arXiv. https://doi.org/10.48550/arXiv.2509.15292

Ding, Y., & Bu, Y. (2025). Political hegemony, imitation isomorphism, and project familiarity: Instrumental variables to understand funding impact on scholar performance. *Quantitative Science Studies*, *6*, 483–504. https://doi.org/10.1162/qss_a_00359

Don't deprioritize curiosity-driven research. (2026). *Nature*, *650*(8102), 524. https://doi.org/10.1038/d41586-026-00469-0

Fahrudin, T. M., Funabiki, N., Brata, K. C., Noprianto, N., Muhaimin, A., & Hindrayani, K. M. (2025). Comparative Analysis of Sentence Transformers for Reference Paper Collection in Five Academic Fields. *Proceedings of the 2025 8th International Conference on Computational Intelligence and Intelligent Systems*, 139–144. https://doi.org/10.1145/3787256.3787277

Gök, A., Rigby, J., & Shapira, P. (2016). The impact of research funding on scientific outputs: Evidence from six smaller E uropean countries. *Journal of the Association for Information Science and Technology*, *67*(3), 715–730. https://doi.org/10.1002/asi.23406

Gusenbauer, M. (2024). Beyond Google Scholar, Scopus, and Web of Science: An evaluation of the backward and forward citation coverage of 59 databases' citation indices. *Research Synthesis Methods*, *15*(5), 802–817. https://doi.org/10.1002/jrsm.1729

IIT-Concerned, Fleming, S. M., Frith, C. D., Goodale, M., Lau, H., LeDoux, J. E., Lee, A. L. F., Michel, M., Owen, A. M., Peters, M. A. K., & Slagter, H. A. (2023). *The Integrated Information Theory of Consciousness as Pseudoscience*. PsyArXiv. https://doi.org/10.31234/osf.io/zsr78

Ioannidis, J. P. A. (2012). Research needs grants, funding and money – missing something? *European Journal of Clinical Investigation*, *42*(4), 349–351. https://doi.org/10.1111/j.1365-2362.2011.02617.x

Jacob, B. A., & Lefgren, L. (2011). The impact of research grant funding on scientific productivity. *Journal of Public Economics*, *95*(9–10), 1168–1177. https://doi.org/10.1016/j.jpubeco.2011.05.005

Kaiser, J. (2026, February 20). NIH research grant funding rates plummeted in 2025. *Science*. https://doi.org/10.1126/science.z1bp5k1

Kamada, S., Ichimura, T., & Watanabe, T. (2015). Recommendation System of Grants-in-Aid for Researchers by using JSPS Keyword. *2015 IEEE 8th International Workshop on Computational Intelligence and Applications (IWCIA)*, 143–148. https://doi.org/10.1109/IWCIA.2015.7449479

Kamada, S., Ichimura, T., & Watanabe, T. (2016). A Recommendation System of Grants to Acquire External Funds. *2016 IEEE 9th International Workshop on Computational Intelligence and Applications (IWCIA)*, 125–130. https://doi.org/10.1109/IWCIA.2016.7805760

Kumar, A., Koley, M., Yegros, A., & Rafols, I. (2024). Priorities of health research in India: Evidence of misalignment between research outputs and disease burden. *Scientometrics*. https://doi.org/10.1007/s11192-024-04980-x

Lane, J. (2009). Assessing the Impact of Science Funding. *Science*, *324*(5932), 1273–1275. https://doi.org/10.1126/science.1175335

Murray, D. L., Morris, D., Lavoie, C., Leavitt, P. R., MacIsaac, H., Masson, M. E. J., & Villard, M.-A. (2016). Bias in Research Grant Evaluation Has Dire Consequences for Small Universities. *PLOS ONE*, *11*(6), e0155876. https://doi.org/10.1371/journal.pone.0155876

Naddaf, M. (2025). Is academic research becoming too competitive? Nature examines the data. *Nature*, *646*(8087), 1036–1037. https://doi.org/10.1038/d41586-025-03119-z

Ostendorff, M., Blume, T., Ruas, T., Gipp, B., & Rehm, G. (2022). Specialized document embeddings for aspect-based similarity of research papers. *Proceedings of the 22nd ACM/IEEE Joint Conference on Digital Libraries*, 1–12. https://doi.org/10.1145/3529372.3530912

Publishers close access to scholarly content such as abstracts due to AI incentives. (2025, September 30). *Access*. https://librarylearningspace.com/publishers-close-access-to-scholarly-content-such-as-abstracts-due-to-ai-incentives/

Rennie, D. (2000). The Contributions of Authors. *JAMA*, *284*(1), 89. https://doi.org/10.1001/jama.284.1.89

Røttingen, J.-A., Regmi, S., Eide, M., Young, A. J., Viergever, R. F., Årdal, C., Guzman, J., Edwards, D., Matlin, S. A., & Terry, R. F. (2013). Mapping of available health research and development data: What's there, what's missing, and what role is there for a global

observatory? *The Lancet*, *382*(9900), 1286–1307. https://doi.org/10.1016/S0140-6736(13)61046-6

Schmutz, A., Salignat, C., Plotkina, D., Devouassoux, A., Lee, T., Arnold, M., Ervik, M., & Kelm, O. (2019). Mapping the Global Cancer Research Funding Landscape. *JNCI Cancer Spectrum*, *3*(4), pkz069. https://doi.org/10.1093/jncics/pkz069

Singh, A., D'Arcy, M., Cohan, A., Downey, D., & Feldman, S. (2022). *SciRepEval: A Multi-Format Benchmark for Scientific Document Representations* (Version 4). arXiv. https://doi.org/10.48550/ARXIV.2211.13308

Thelwall, M., Simrick, S., Viney, I., & Van Den Besselaar, P. (2023). What is research funding, how does it influence research, and how is it recorded? Key dimensions of variation. *Scientometrics*, *128*(11), 6085–6106. https://doi.org/10.1007/s11192-023-04836-w

Tian, W., Cai, R., Fang, Z., Xie, Q., Hu, Z., & Wang, X. (2024). Research funding in different SCI disciplines: A comparison analysis based on Web of Science. *Quantitative Science Studies*, *5*(3), 757–777. https://doi.org/10.1162/qss_a_00315

Ubfal, D., & Maffioli, A. (2011). The impact of funding on research collaboration: Evidence from a developing country. *Research Policy*, *40*(9), 1269–1279. https://doi.org/10.1016/j.respol.2011.05.023

Wolff, B., Seidlmayer, E., & Förstner, K. U. (2024). *Enriched BERT Embeddings for Scholarly Publication Classification*. https://doi.org/10.1007/978-3-031-65794-8_16

Yegros-Yegros, A., Van De Klippe, W., Abad-Garcia, M. F., & Rafols, I. (2020). Exploring why global health needs are unmet by research efforts: The potential influences of geography, industry and publication incentives. *Health Research Policy and Systems*, *18*(1), 47. https://doi.org/10.1186/s12961-020-00560-6

Yin, Y., Dong, Y., Wang, K., Wang, D., & Jones, B. F. (2022). Public use and public funding of science. *Nature Human Behaviour*, *6*(10), 1344–1350. https://doi.org/10.1038/s41562-022-01397-5

Zhang, Y., Chen, Q., Yang, Z., Lin, H., & Lu, Z. (2019). BioWordVec, improving biomedical word embeddings with subword information and MeSH. *Scientific Data*, *6*(1), 52. https://doi.org/10.1038/s41597-019-0055-0

Zhang, Z., Patra, B. G., Yaseen, A., Zhu, J., Sabharwal, R., Roberts, K., Cao, T., & Wu, H. (2023). Scholarly recommendation systems: A literature survey. *Knowledge and Information Systems*, *65*(11), 4433–4478. https://doi.org/10.1007/s10115-023-01901-x

Zhang, Z., Yaseen, A., & Wu, H. (2024). Scholarly recommendation system for NIH funded grants based on biomedical word embedding models. *Natural Language Processing Journal*, *8*, 100095. https://doi.org/10.1016/j.nlp.2024.100095

Zhu, J., Patra, B. G., Wu, H., & Yaseen, A. (2023). A novel NIH research grant recommender using BERT. *PLOS ONE*, *18*(1), e0278636. https://doi.org/10.1371/journal.pone.0278636